\documentclass[manuscript]{acmart}

\AtBeginDocument{%
  \providecommand\BibTeX{{%
    \normalfont B\kern-0.5em{\scshape i\kern-0.25em b}\kern-0.8em\TeX}}}

\setcopyright{acmcopyright}
\copyrightyear{2023}
\acmYear{2023}
\acmDOI{XXXXXXX.XXXXXXX}

\acmConference[pluralism@CSCW '23]{Many Worlds of Ethics: Ethical Pluralism in CSCW}{October 14, 2023}{Minneapolis, MN}

\begin{document}

\title{Towards Incorporating Researcher Safety into Information Integrity Research Ethics}

\author{Joseph S. Schafer}
\email{schaferj@uw.edu}
\orcid{0000-0002-6921-2074}
\affiliation{%
  \institution{University of Washington Center for an Informed Public}
  \city{Seattle}
  \state{Washington}
  \country{USA}
  \postcode{98195}}
  \affiliation{%
  \institution{University of Washington Human-Centered Design \& Engineering}
  \city{Seattle}
  \state{Washington}
  \country{USA}
  \postcode{98195}}
\author{Kate Starbird}
\email{kstarbi@uw.edu}
\orcid{0000-0003-1661-4608}
\affiliation{%
  \institution{University of Washington Center for an Informed Public}
  \city{Seattle}
  \state{Washington}
  \country{USA}
  \postcode{98195}}
  \affiliation{%
  \institution{University of Washington Human-Centered Design \& Engineering}
  \city{Seattle}
  \state{Washington}
  \country{USA}
  \postcode{98195}}

\renewcommand{\shortauthors}{Schafer \& Starbird}

\begin{abstract}
 Traditional research ethics has mainly and rightly been focused on making sure that participants are treated safely, justly, and ethically, to avoid the violation of their rights or putting participants in harm’s way. Information integrity research within CSCW has also correspondingly mainly focused on these issues, and the focus of internet research ethics has primarily focused on increasing protections of participant data. However, as branches of internet research focus on more fraught contexts such as information integrity and problematic information, more explicit consideration of other ethical frames and subjects is warranted. In this workshop paper, we argue that researcher protections should be more explicitly considered and acknowledged in these studies, and should be considered alongside more standard ethical considerations for participants and for broader society. 
\end{abstract}

\begin{CCSXML}
<ccs2012>
   <concept>
       <concept_id>10003120.10003130.10003233.10010519</concept_id>
       <concept_desc>Human-centered computing~Social networking sites</concept_desc>
       <concept_significance>500</concept_significance>
       </concept>
   <concept>
       <concept_id>10003120.10003130.10003131.10011761</concept_id>
       <concept_desc>Human-centered computing~Social media</concept_desc>
       <concept_significance>500</concept_significance>
       </concept>
   <concept>
       <concept_id>10003120.10003130.10003134</concept_id>
       <concept_desc>Human-centered computing~Collaborative and social computing design and evaluation methods</concept_desc>
       <concept_significance>500</concept_significance>
       </concept>
 </ccs2012>
\end{CCSXML}

\ccsdesc[500]{Human-centered computing~Social networking sites}
\ccsdesc[500]{Human-centered computing~Social media}
\ccsdesc[500]{Human-centered computing~Collaborative and social computing design and evaluation methods}

\keywords{information integrity, misinformation, disinformation, research ethics, researcher safety}

\received{15 August 2023}
\received[Revised]{24 September 2023}

\maketitle

\section{Introduction}
Research into information integrity, such as misinformation, disinformation, rumors, and other problematic information, has become increasingly prominent  as this topic impacts a wide array of issues, including climate change \cite{treen_online_2020}, public health \cite{koltai_vaccine_2020}, and elections \cite{abilov_voterfraud2020_2021, akbar_political_2022}.  Since much of this work focuses on the spread of misinformation and disinformation within online communities and on social media platforms, the CSCW research community has published significant research in this area and continues to prioritize research into these topics. This research, when being ethically evaluated, has most often been judged in the light of maturing ideas of internet research ethics, particularly around the ethics of researching nominally ‘public’ data on the internet which users perceive as private, in conversations adjacent to the CSCW community \cite{fiesler_participant_2018, hudson_go_2004, zimmer_but_2010}. However, the research climate for information integrity research, as well as other broad topics of research into related harmful or hate-based online behaviors relevant to CSCW scholarship, cause researchers to face significant risks when conducting this work. This can include online or offline harassment and threats against researchers for doing this work, mental health costs to researchers working on these stressful topics, or legal actions threatened or taken against researchers, some of which we have previously outlined in collaborative work with Daniela K. Rosner \cite{schafer_participatory_2023}. It's worth noting that risks like this are not unique to information integrity researchers, as other groups have been recently targeted due to the spread of misinformation and disinformation, such as election workers facing threats \cite{zakrzewski_election_2022} and medical scientists facing social media backlash \cite{nolleke_chilling_2023}. These risks show that additional ethical frameworks to protect researchers are necessary to incorporate in our studies, which have been explored in other research fields and should be more deeply considered within the CSCW community. In the remainder of this paper, we will outline some work in these fields which are relevant to the CSCW community’s research into information integrity.

\section{Prior Work on Researcher Safety}

Explicitly considering harm to researchers in research ethics is not a novel concept. Christian Fuchs and Graham Meikle argue in a 2018 paper for the need for a ‘critical-realist’ perspective to internet research ethics between the two approaches of obtaining informed consent for every social media post, or assuming every post is fully public and usable for research \cite{fuchs_dear_2018}. Fuchs and Meikle’s paper is focused on online “ideology critique” research, which is a separate focus from information integrity research, but invokes many of the same concerns. Fuchs and Meikle also explicitly acknowledge that typical privacy-protecting data practices like asking for informed consent for participants means that “The researcher may be next in line for being harassed or threatened” \cite{fuchs_dear_2018}. Adrienne L. Massanari also describes the challenges of ethically conducting research into online alt-right groups, arguing that these risks from participants to researchers represent a new ‘gaze’ on research, and needs to account for risks to the researcher beyond the immediate context of the study, across both time and the researchers’ relationships to family and friends \cite{massanari_rethinking_2018}. Massanari specifically recommends engagement with approaches that improve researcher solidarity and networks of support, and building better institutional support for researchers working on research into alt-right groups \cite{massanari_rethinking_2018}. 

Brit Kelley and Stephanie Weaver describe doing ethnographic research in online spaces after Gamergate, and particularly focusing on vulnerabilities potentially existing in both directions between researchers and participants \cite{kelley_researching_2020}. They also make explicit that participants can be hostile or threatening toward researchers in general, as well as specific researchers whose identities are denigrated by these groups, such as researchers of color studying online white nationalists. In our own prior work, we have also explored related ideas of the dangers of reciprocity complicating research ethics on these topics, looking at how to ethically and effectively use participatory design methods for misinformation, disinformation, and online hate research \cite{schafer_participatory_2023}, alongside the need for researchers to consider the degree to which authentic representation of participants and real-world impacts can be safely achieved and sought in research. Ahmer Arif and colleagues also address some challenges to researchers when studying Russian information operations on Twitter, describing how these information operations "are effective at what many have argued they intend to do—sowing doubt, creating confusion" \cite{arif_acting_2018}. Ahmer Arif's dissertation further articulates these challenges, writing that "studying disinformation can attract harassment and unwanted attention" and that "my colleagues and I have had to pay attention to our psychological and physical safety to do parts of this work" \cite{arif_troubling_2020}.

Similar concerns have been discussed in using social media for other aspects of the research process, such as in the work of Kari Dee Vallury and colleagues describing the challenges of safely using social media for research recruitment \cite{vallury_going_2021}. While recruiting participants for an abortion-related study, the researchers were targeted by other social media users, for which they received institutional support but in a somewhat patchy, unprepared process \cite{vallury_going_2021}. The authors argue that institutions need to have clearer policies and guidelines that can be used for future cases of harassment like this, to make this experience not as harmful for researchers \cite{vallury_going_2021}. Researchers in feminist media studies and other fields have also talked about the harassment they have faced throughout the research process, such as Shira Chess and Adrienne Shaw’s case study of the negative reactions to their (and others’) research into Gamergate \cite{chess_conspiracy_2015}, F. Vera-Gray’s experience of harassment when men’s rights activists found her recruitment materials for a study on street harassment \cite{vera-gray_talk_2017}, and the CSCW workshop in 2020 on challenges associated with social media and public scholarship organized by Sarah A. Gilbert and colleagues \cite{gilbert_public_2020}.

\section{Lessons for Information Integrity Research}
In light of this prior research, several key lessons are important to consider for CSCW research on information integrity. First, we echo the calls from Kari Dee Vallury et. al. \cite{vallury_going_2021} and Adrienne Massanari \cite{massanari_rethinking_2018} that  institutional and collaborative networks of support need to be built. CSCW researchers studying information integrity should take part in creating these systems, both due to the field’s contributions to knowledge of these phenomena and its knowledge of technology-incorporated group organizing broadly. While we cannot prevent all of the challenges and risks of this research, creating networks of care and support could help to mitigate these risks. 

These papers also make explicit the need to consider researchers as a worthy group of stakeholders for ethical consideration, as well as recognizing that not all researchers face these risks and harms equally. The positionality of researchers conducting projects will influence the degree to which researchers are likely to face these threats, as well as how impactful these harms will be. Concerns for ethical research should consider the safety needs of both participants and the research team.

Acknowledging the safety of researchers in information integrity research studies could also help to address other criticisms recently directed at these fields, by helping to foreground research positionality. Multiple recent credible criticisms of the field of ‘disinformation studies’ have surfaced, some of which are described in the ‘After Disinformation Studies’ volume from the University of North Carolina’s Center for Information, Technology, and Public Life and its introduction essay by Theophilé Lenoir and Chris Anderson \cite{lenoir_introduction_2023}. Studying potentially riskier questions such as the limits of rational models of misinformation susceptibility, and how the field of information integrity has “lacked analyses of power and interest” \cite{lenoir_introduction_2023}, could potentially be better conducted if concerns about researcher safety and protection in study design were foregrounded. While this would not address all of these criticisms, this could help as a starting point to begin to address these gaps in the field.

This prior work also points in some promising directions for future work on internet research ethics in the field of information integrity. For example, researchers could conduct an overview of existing research practices in these fields to understand what existing ethical considerations are widespread within the field, potentially through doing systematic literature reviews for ethical practices or through interviewing other researchers in these fields. CSCW scholars could also bring these works more directly into conversation with prior CSCW-adjacent literature on online research participant privacy, and work towards more practical, contextual guidelines for conducting research on these topics. Additionally, using participatory methods to design researcher support networks such as Vallury et. al. \cite{vallury_going_2021} and Massanari \cite{massanari_rethinking_2018} describe for information integrity researchers could both improve conditions for researchers in these fields and inform future similar mutual support group designs. These efforts could also build off of broader efforts within CSCW to improve systems for science communication, such as the work of Spencer Williams and colleagues \cite{williams_an_2022}, as researcher safety in science communication is essential.

We must also note that the rapidly changing online landscape alters the vulnerabilities and threat models for researchers working in these fields. Recent and impending shifts in data and API access on popular platforms for research like Twitter \cite{calma_scientists_2023}, Reddit \cite{mehta_reddit_2023} \footnote{Reddit’s API changes primarily impact third-party developers, not researchers, as they still have a new portal for researcher API access \cite{reddit_reddit_2023}. However, Pushshift, a third-party tool commonly used in Reddit research, was banned around the same time as the API changes \cite{lift_ticket83_reddit_2023}.}, and Facebook and Instagram \cite{lawler_meta_2022}, and increasing user migrations to more dispersed and less obviously public social media systems like Mastodon, Discord, and Bluesky, should give information integrity researchers cause to reflect on how to effectively and ethically study platforms, in ways that are protective of both participants and of researchers. Much of the previous work on researcher safety focused on internet landscapes prior to these shifts; understanding what changes these shifts cause for ethical internet research is of critical importance. 

\section{Discussion prompts}
At the "Many Worlds of Ethics: Ethical Pluralism in CSCW" workshop, we hope to use this workshop paper to start dialogues around these areas. In particular, we would like to discuss the following questions:
\begin{itemize}
    \item What considerations should CSCW researchers use to think about their own safety and those of their research teams?
    \item What supports can the broader CSCW community build to make this work sustainable?
    \item What other lenses on researcher safety would be useful to bring into CSCW work on information integrity, and what could CSCW reflect back to these fields? 
\end{itemize}

\section{Conclusion}
Information integrity research should more explicitly consider the safety of researchers themselves as part of research ethics. As other fields, including feminist media studies, have previously explored, work in information integrity and related areas can result in researchers taking on significant risks. Creating effective researcher support systems and designing studies which prioritize researcher safety are ways that researchers in these fields could be better protected from these harms.

\begin{acks}
We are grateful for support from the National Science Foundation, from NSF Awards \#1749815 and DGE-2140004. We are also supported in part by the University of Washington Center for an Informed Public, the John S. and James L. Knight Foundation (G-2019-58788), the Election Trust Initiative, and the William and Flora Hewlett Foundation (2022-00952-GRA). Any opinions, findings, and conclusions or recommendations expressed in this material are those of the authors and do not necessarily reflect the views of the National Science Foundation or other funders.
\end{acks}

\bibliographystyle{ACM-Reference-Format}
\bibliography{pluralism}

\end{document}